\documentclass[times,twocolumn,final]{elsarticle}

\usepackage{framed,multirow}
\usepackage{todonotes}
\usepackage{amssymb}
\usepackage{latexsym}
\usepackage{subcaption}
\usepackage{float}
\usepackage{amsmath}

\usepackage{url}
\usepackage{xcolor}
\newsavebox{\bigimage}

\usepackage{hyperref}

\definecolor{newcolor}{rgb}{.8,.349,.1}

\begin{document}
\begin{frontmatter}

\title{Towards the use of multiple ROIs for radiomics-based survival modelling: finding a strategy of aggregating lesions}

\author[1,2]{Agata Ma\l gorzata Wilk}
 \ead{agata.wilk@polsl.pl}
\author[1]{Andrzej Swierniak}
\author[3]{Andrea d'Amico}
\author[4]{Rafa\l ~ Suwi\'{n}ski}
\author[1]{Krzysztof Fujarewicz}
\author[1]{Damian Borys\corref{cor1}}
\ead{damian.borys@polsl.pl}
\cortext[cor1]{Corresponding author: 
  Tel.: +48-32-237-11-59;  
  fax: +48-32-237-16-55;}

\address[1]{Department of Systems Biology and Engineering, Silesian University of Technology, Akademicka 16, Gliwice 44-100, Poland}
\address[2]{Department of Biostatistics and Bioinformatics, Maria Sklodowska-Curie National Research Institute of Oncology, Gliwice Branch, Wybrzeze AK 15, Gliwice 44-102, Poland}
\address[3]{Department of Nuclear Medicine and Endocrine Oncology, PET Diagnostics Unit, Maria Sklodowska-Curie National Research Institute of Oncology, Gliwice Branch, Wybrzeze AK 15, Gliwice 44-102, Poland}
\address[4]{II-nd Radiotherapy and Chemotherapy Clinic, Maria Sklodowska-Curie National Research Institute of Oncology, Gliwice Branch, Wybrzeze AK 15, Gliwice 44-102, Poland}

\begin{abstract}
%%%
\textit{Background}. Radiomic features, derived from a region of interest (ROI) in medical images, are valuable as prognostic factors. Selecting an appropriate ROI is critical, and many recent studies have focused on leveraging multiple ROIs by segmenting analogous regions across patients --— such as the primary tumour and peritumoral area or subregions of the tumour. These can be straightforwardly incorporated into models as additional features. However, a more complex scenario arises for example in a regionally disseminated disease, when multiple distinct lesions are present. 

\textit{Aim}. This study aims to evaluate the feasibility of integrating radiomic data from multiple lesions into survival models. We explore strategies for incorporating these ROIs and hypothesise that including all available lesions can improve model performance.

\textit{Methods}. While each lesion produces a feature vector, the desired result is a unified prediction. We propose methods to aggregate either the feature vectors to form a representative ROI or the modeling results to compute a consolidated risk score. As a proof of concept, we apply these strategies to predict distant metastasis risk in a cohort of 115 non-small cell lung cancer patients, 60\% of whom exhibit regionally advanced disease. Two feature sets (radiomics extracted from PET and PET interpolated to CT resolution) are tested across various survival models using a Monte Carlo Cross-Validation framework.

\textit{Results}. Across both feature sets, incorporating all available lesions --- rather than limiting analysis to the primary tumour --- consistently improved the c-index, irrespective of the survival model used.

\textit{Conclusion}. Lesions beyond the primary tumour carry information that should be utilised in radiomics-based models to enhance predictive ability.

\end{abstract}

\begin{keyword}

multiple ROIs \sep radiomics \sep survival models \sep ROI aggregation \sep risk aggregation
\end{keyword}
\end{frontmatter}

%\linenumbers

\section{Introduction}
Non-invasive, quick to acquire, and accurate imaging is a staple of modern medicine. With various available modalities, it allows for observing the structural, metabolic, and biochemical state of the patient's body, proving invaluable for diagnostics, therapy planning and management. In addition to their standard application based on radiological asessment, resulting digital images can be treated as data used in machine learning models \cite{Gillies2016}. Although deep learning can be applied directly to raw images, a major obstacle is the cohort size necessary to construct a well-calibrated model \cite{Chen2022B}. A compromise solution is radiomics, where numerical features are first extracted from a specific part of the image.

% These can be fed into statistical models, achieving better performance compared to clinical or radiological features \cite{Park2021}. The growing interest in the field, related among others to oncology \cite{TAGLIAFICO2020,Chetan2020} can largely be attributed to consistent efforts to make the extracted features robust and reproducible \cite{Zwanenburg2020}.

One of the challenges in radiomics is the definition of the region of interest (ROI). Typically, in cancer-related studies, an intuitive region is the primary tumour itself, however, various other ROI delineations have been explored. In \cite{Wang2022}, radiomic features were extracted from intra-tumour region as well as the peritumoural area and applied for survival risk prediction in early-stage lung cancer. Shan et al. \cite{Shan2019} used ROIs corresponding to the lesion and the peritumoural area to predict recurrence in hepatocellular carcinoma. In \cite{Lu2021}, a dual-region radiomics model incorporating both primary tumor and lymph node computed tomography features achieved improved survival prediction in esophageal squamous cell cancer compared to a single-region model. Chen et al. \cite{Chen2022A} described a deep learning model for MGMT promoter methylation prediction using radiomic features extracted from the whole tumour region and the tumour core. Xu et al. \cite{Xu2022} presented a diagnostic approach utilizing five different ROIs in two types of breast ultrasound images, including the whole tumour region, the strongest perfusion region, and the surrounding region. In \cite{Feng2022}, radiation-induced skin toxicity was investigated using six types of ROIs based on radiation doses. Chen et al. \cite{Chen2018} predicted metastasis in rectal cancer based on ROIs delineating tumour area, peritumoural fat, and the largest pelvic lymph node.  In \cite{Zhou2020}, three ROIs of different sizes created around the nodule center served for differentiation between benign and malignant thyroid nodules. Han et al. \cite{Han2020} used features extracted from the tumour zone and the tumour-liver interface to predict predominant histopathological growth patterns of colorectal liver metastases. In \cite{Wang2024}, axillary lymph node (ALN) metastasis status in breast cancer was predicted using radiomics extracted from tumour and ALN ROIs in MRI and mammography images. Dammak et al. \cite{Dammak2024} distinguished between lung cancer recurrence and benign radiation-induced injury based on six semi-automatically contoured ROIs (each initialised with a RECIST line drawn by a specialist). In \cite{Peng2024}, immunotherapy response in non-small cell lung cancer was predicted based on radiomic features extracted from three sub-regions of the primary tumour obtained through k-means clustering. Hou et al. \cite{Hou2024} constructed a model predicting neutrophil-lymphocite ratio for lung cancer patients based on CT radiomics from five anatomical regions. A work by Zhang et al. \cite{Zhang2023} demonstrates the influence of peritumoural margin on the performance of radiomics models. A later study \cite{Zhang2024} describes a model combining deep learning and conventional radiomics for ROIs comprising the tumour region and peritumoural area.  Similarly, in \cite{Zhang2021}, radiomic and deep features are integrated to classify breast lesions. In \cite{Nie2024}, ROIs resulting from radiotherapy planning are considered --- gross tumour volume, planning tumour volume, and the difference between the two. Even in studies focusing on extraction of deep features, multiple ROI delineations are used, such as in \cite{Wang2021}, where the VOI used to predict nodal metastasis in lung cancer was divided into sections representing tumour core and peritumoral area. These studies show that including other ROIs can improve the predictive ability. Still, their inclusion in the model remains relatively straightforward since they can be determined for each of the patients and treated as additional features, introducing no ambiguity. 

However, in many cases the tumour is already locally spread \cite{SEER}, and in addition to the primary tumour there are also secondary lesions and involved lymph nodes. All these structures may constitute separate regions of interest and carry information valuable for prediction. While many works, as mentioned above, present models including different (often multiple) \textit{definitions} of the region of interest, the problem of using actually distinct regions, whose number is inconsistent between patients, is addressed much less frequently. Notably, Zhao et al. \cite{Zhao2022} proposed a method based on "meta histograms" and applied it for PET/CT radiomics in lung adenocarcinoma to train classification models corresponding to 3- and 4-year overall survival, as well as tumour grade and risk. The core idea of this method is aggregation of multiple ROIs to form a representative feature vector. However, in the article \cite{Kurczyk2020} dedicated to mass spectrometry imaging, a different strategy was described. It was discussed that a "single-pixel" approach (using all available data for training the model and aggregating the results) allowed for better classification quality than an aggregated spectrum approach. 

Expanding on these ideas, we explore the feasibility of using aggregation of ROIs or risk scores to integrate multiple distinct lesions in radiomics-based survival models. We formulate various aggregation methods and we hypothesise that incorporating all lesions may allow for a more accurate prediction than for the primary tumour alone. Since no datasets containing information about multiple lesions are available to use for a systematic comparison, we compiled a sample dataset. We chose lung cancer, because it is often diagnosed at a late, locally advanced stage (and thus involves a high probability of multiple ROIs being present), and it has already been shown  \cite{Wilk2022,Wilk2023,Fujarewicz2022,Coroller2015} that radiomic features extracted from the primary tumour have the potential for predicting the risk of metastatic dissemination. We demonstrate the comparison of c-indices for the "meta histogram" strategy as well as original ROI or risk aggregation methods for several different classes of survival models.

\section{Materials and methods}
The retrospectively collected study cohort, described in detail elsewhere \cite{Wilk2023}, consists of 115 Polish non-small cell lung cancer patients treated in the Maria Sklodowska-Curie National Research Institute of Oncology Gliwice Branch. 
In short, data for over 800 lung cancer patients treated in the Maria Sklodowska-Curie National Research Institute of Oncology Gliwice Branch were retrospectively retrieved from the hospital system. We focused on non-small cell lung cancer (NSCLC) as the most prevalent subtype, excluding 115 cases of other subtypes. As the group most likely to benefit from accurate metastasis risk assessment, we selected patients treated with curative intent using chemoradiotherapy. 132 of them had planning PET/CT images available. 17 cases with incomplete clinical data or distant metastases detectable at diagnosis were also excluded. 
The study was approved by the institutional Ethics Committee (KB/430-48/23) and the data were anonymized before the analysis. 

The endpoint considered in this study was metastasis-free survival (MFS), defined as the time from diagnosis to detection of distant metastases (event) or the last screening without detectable distant metastases (censored observations). It is important to note, that as implied by this definition, censoring from the MFS point of view does not coincide with the patient's death.  

\subsection{PET/CT image acquisition and processing}

As part of radiotherapy planning, PET/CT images were acquired using Philips GeminiGXL 16 (Philips, Amsterdam, Netherlands) (24 patients) and Siemens Biograph mCT 131 (Siemens
AG, Munich, Germany) (91 patients). An experienced nuclear medicine expert contoured all the radiologically changed regions within the lung. Taking into consideration the validity of texture features, we excluded regions smaller than two voxels. Using PyRadiomics v3.1.0 \cite{vanGriethuysen2017} with bin width set to 0.1, we extracted 100 radiomic features for each region of interest (ROI). Before calculating the features we standardized the PET images using body weight (SUVbw) and used the images both in the original resolution (PET dataset) and interpolated to the CT resolution with the nearest neighbour algorithm (PET\_CT dataset). None of the additional filters were used in the pre-processing step.

\subsection{Inter-ROI heterogeneity}

To assess whether the number of ROIs is related to MFS, we divided the patients into three groups of similar cardinalities --- with only one contoured ROI, with two to three ROIs, and with four or more ROIs. We compared the Kaplan-Meier curves for the resulting groups using the log-rank test. 

To assess the level of inter-ROI diversity, several concerns had to be addressed. Firstly, some of the radiomic features can take negative values, which excludes certain popular dissimilarity measures, in particular entropy-based indices. In addition, the number of ROIs was not consistent between patients. We decided to employ several distance- and correlation-based indices, listed below:
% \begin{itemize}
%     \item Canberra distance,
%     \item Euclidean distance,
%     \item Minkowski distance,
%     \item Kendall distance,
%     \item Spearman distance.
% \end{itemize}

% Let $x = [x_1,x_2,...,x_n]$ and $y = [y_1,y_2,...,y_n]$ be numerical vectors in an $n$-dimensional vector space of real numbers.

\noindent \textbf{Canberra distance}.  The Canberra distance $d_C(x,y)$ is defined as:
\begin{equation}
    d_C(x,y) = \sum^n_{i=1} \frac{|x_i-y_i|}{|x_i|+|y_i|}
\end{equation}
\textbf{Euclidean distance}.  The Euclidean distance $d_E(x,y)$ is defined as:
\begin{equation}
    d_E(x,y) = \sqrt{\sum^n_{i=1} (x_i-y_i)^2}
\end{equation}
\textbf{Minkowski distance}.  The Minkowski distance $d_M(x,y)$ is defined as:
\begin{equation}
    d_M(x,y) = \left(\sum^n_{i=1} |x_i-y_i|^p\right)^{\frac{1}{p}}
\end{equation}
\textbf{Kendall distance}.  The Kendall distance $d_K(x,y)$, based on the Kendall correlation coefficient, is defined as:
\begin{equation}
    d_K(x,y) = 1- |\tau| = 1- \left|\frac{2}{n(n-1)}\sum_{i<j} {sgn}(x_{i}-x_{j})\operatorname {sgn}(y_{i}-y_{j})\right|
\end{equation}
\textbf{Spearman distance}.  The Spearman distance $d_S(x,y)$, based on the Spearman correlation coefficient, is defined as:
\begin{equation}
    d_S(x,y) = 1- |\rho| = 1-\left| \frac{cov(R(x), R(y))}{\sigma_{R(x)} \sigma_{R(y)}} \right|
\end{equation}
where $R(x)$ and $R(y)$ are $x$ and $y$ converted to rank vectors. Due to the vast differences in magnitude levels of the radiomic features, Pearson's correlation coefficient was impractical for this application. 

For a particular patient, the heterogeneity index is equal to $0$ if there is only one ROI, and to the average value for all distinct pairs of ROIs otherwise. For consistency against the number of ROIs, the patients were again stratified into three groups, according to terciles of the heterogeneity index values. Naturally, in our cohort, the one-ROI group and the low-heterogeneity group for any given index are equal. 

\subsection{Methods of handling multiple ROIs}

\begin{figure*}[!t]
\centering
\includegraphics[scale=.6]{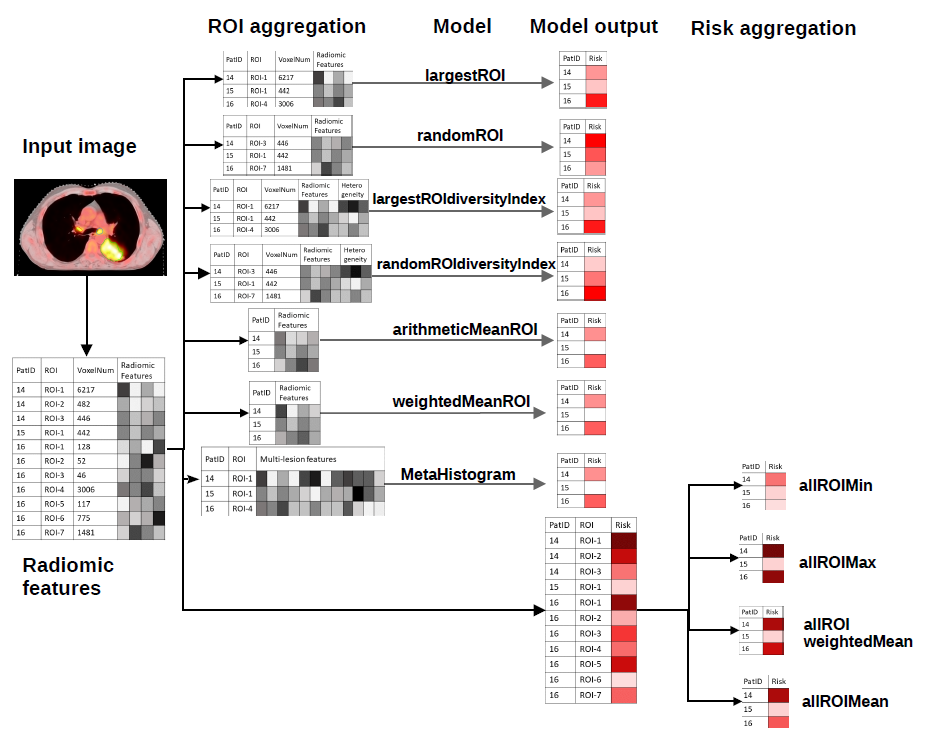}
\caption{Methods of handling multiple ROIs. Radiomic features (represented by grey rectangles) are extracted from the input image for each region of interest. In the ROI aggregation methods, a representative ROI is constructed. The ROIs serve as input for a survival model, whose output is a risk score (represented by red rectangles). In risk aggregation methods, risk scores are aggregated to obtain a single value for every patient. See a more detailed description in the text.}
\label{fig:ROImethods}
\end{figure*}

Generally, survival-type models take as input a collection of observations (patients, objects), each of them consisting of a feature vector, in our case radiomic features for the ROI, and the response, which is a pair of values denoting time-to-event and status (event or censoring). After fitting, the model can be used for new observations, assigning a risk value corresponding to a new feature vector. Utilizing multiple feature vectors, non-equivalent between patients, requires aggregation of either ROIs before the model training or risks thereafter. A schematic illustration of the methods is shown in Figure \ref{fig:ROImethods}. They are listed and described in the following Section.

\subsubsection{Aggregation of ROIs}
The main idea of ROI aggregation approaches is to select or construct a single ROI to represent each patient. The dataset consists of fewer data points, but the models can be used as is, and the results require no further processing. 

\textit{largestROI}. In the case of medical imaging, the ROIs can be intuitively ordered according to their size, that is the number of voxels. The largest one, usually corresponding to the primary tumour, is a default choice for the modelling, and as such will be treated as the reference method. Although it is a straightforward approach, its major disadvantage is the loss of information from the remaining ROIs. 

\textit{randomROI}. For each patient, a random ROI is selected. Since the result is largely dependent on chance, it may differ for each realization of the method. Taking into account considerable variability between ROIs for a single patient, the instability it introduces in the model makes it practically unsuitable for clinical application.  Nevertheless, this method is useful when there exists no intrinsic ordering of the measurements. 

\textit{largestROIdiversityIndex}. For each patient, the heterogeneity indices described above are calculated based on all the ROIs. The five resulting variables are concatenated to the radiomic feature vector from the largest ROI, and can subsequently be selected within the model. This method utilizes, to some extent, information from all the available data points.

\textit{randomROIdiversityIndex}. Similar to \textit{largestROIdiversityIndex}, except diversity indices are considered together with the radiomic features extracted from a randomly selected ROI.

\textit{arithmeticMeanROI}. A representative ROI $\hat{x}$ is constructed as the arithmetic average of all corresponding ROIs. 
Although it uses information from all contoured regions, in this method many radiomic features, particularly shape features, lose their interpretability. 

\textit{weightedMeanROI}. The representative ROI is calculated as the weighted mean of all regions scaled by the volumes of the ROIs. 

\textit{MetaHistogram}. The method proposed by Zhao et al.\cite{Zhao2022}. Briefly, for each patient a "meta histogram" is constructed --- the "bins" are values of a particular radiomic feature for ROIs arranged in descending order of size. Next, multi-lesion features are extracted as characteristics of the "meta histogram": mean, variance, sum, skewness, kurtosis, energy, and entropy. Since this approach results in a large number of features, correlation-based redundancy filtering was first employed with a 0.9 threshold.

\subsubsection{Aggregation of risks}

In the following methods, all available patient ROIs are used for fitting. When the trained model is applied for prediction, each ROI is assigned its own risk, resulting in multiple scores per patient. Only then, aggregation is performed on the risk values. Since all ROIs are included in the dataset, more data points are available for training. However, high levels of intra-patient heterogeneity can contribute to noise in the data.

\textit{allROIMin}. The smallest risk is chosen for each patient.

\textit{allROIMax}. Opposite to the previous method,  the highest risk is chosen.

\textit{allROIWeightedMean}.The risk for the patient is an average of all the risks for the corresponding ROIs weighted with ROI volumes.

\textit{allROIMean}. Analogous to the previous method, but the patient's risk is calculated as an arithmetic average of ROI risks.

\subsection{Survival models}
Survival models are used in a setting where the response variable consists of pairs $(t_i,\delta_i)$, where $\delta_i$ represents the censoring status, and $t_i$ is the time to event if $\delta_i=1$ or time to last follow-up if $\delta_i=0$. Typically, the output of such models can be interpreted as a relative risk --- the higher the risk score, the sooner the event should occur. Several survival models, representing different classes, usually coupled with some variable selection strategy, were included in the analysis.

\textit{CoxStepAIC}. The classic Cox proportional hazards regression, with forward stepwise selection of variables according to the Akaike Information Criterion (AIC). Due to its popularity in survival analysis, it is treated as the reference model.

\textit{Coxnet}. Regularized Cox regression, with embedded feature selection. The optimal $\lambda$ value was determined in cross-validation. The R package \textit{glmnet} was used \cite{Friedman2010,Simon2011}.

\textit{Weibull}.  Model-based boosting (using \textit{mboost} package) with Weibull accelerated failure time (AFT) model \cite{Hofner2014,mboost}.

\textit{Loglog}.  Model-based boosting with Loglog AFT model.

\textit{Lognormal}.  Model-based boosting with Lognormal AFT model.

\textit{randomForest}. A random survival forest model grown with 1000 trees. The package \textit{RandomForestSRC} was used \cite{randomforestsrc,Ishwaran2008}.

\textit{SVMregression}. A survival support vector machine \cite{VanBelle2010} with a regression model and an additive kernel function. The package \textit{survivalsvm} was used \cite{survivalsvm}.

\textit{SVMvanbelle1}. A survival support vector machine with a Van Belle model and an additive kernel function.

\subsection{Model quality evaluation}
Since for the risk aggregation methods the dataset contained multiple ROIs from the same patient, the validation scheme had to be specifically constructed to prevent information leakage. This would be a situation when ROIs from the same patient were placed in the training set and test set in the same iteration. To avoid it, and to ensure unbiased comparisons between models and aggregation methods, a tailored Monte Carlo cross-validation partitioning was constructed, and fixed for all the schemes. 

In each of 1000 iterations, the patients were partitioned into the training group (2/3) and the test group (1/3). Then, depending on the ROI handling method, the training and test sets were constructed to contain either the representative ROIs for all patients from the corresponding group or all ROIs for patients from the corresponding group. Thus, in every iteration and scheme, the same patient groups were used for training and testing, albeit in the risk aggregation models the actual sizes of the training and test sets differ.

\section{Statistical analysis} 

We employed Harrel's c-Index to evaluate the predictive ability of the models \cite{Harrell1996}. Defined as a ratio of concordant observation pairs (where higher risk corresponds to shorter time-to-event) to all comparable pairs, it can be viewed as a survival analysis counterpart of the area under the ROC curve (AUC) in classification. Indeed, the value 0.5 indicates a completely random model, while 1 is a perfect model. Because of the large sample sizes (1000 iterations constituting a group), we used Cohen's $d$ effect size as an indicator of the strength of the difference against the reference scheme. The standard interpretation was assumed, with $|d|<0.2$ treated as a negligible effect, $0.2\leq|d|<0.5$ meaning small effect, $0.5\leq|d|<0.8$ medium effect and otherwise large effect.

For comparison of survival between two or more groups, we used the log-rank test, assuming the standard significance threshold $p=0.05$.

All statistical analyses were conducted using the R environment for statistical computing, version 4.1.3  (R Core Team, 2022).
\section{Results}
\subsection{Patient characteristics}

\begin{figure}[!t]
\centering
\includegraphics[width=.45\textwidth]{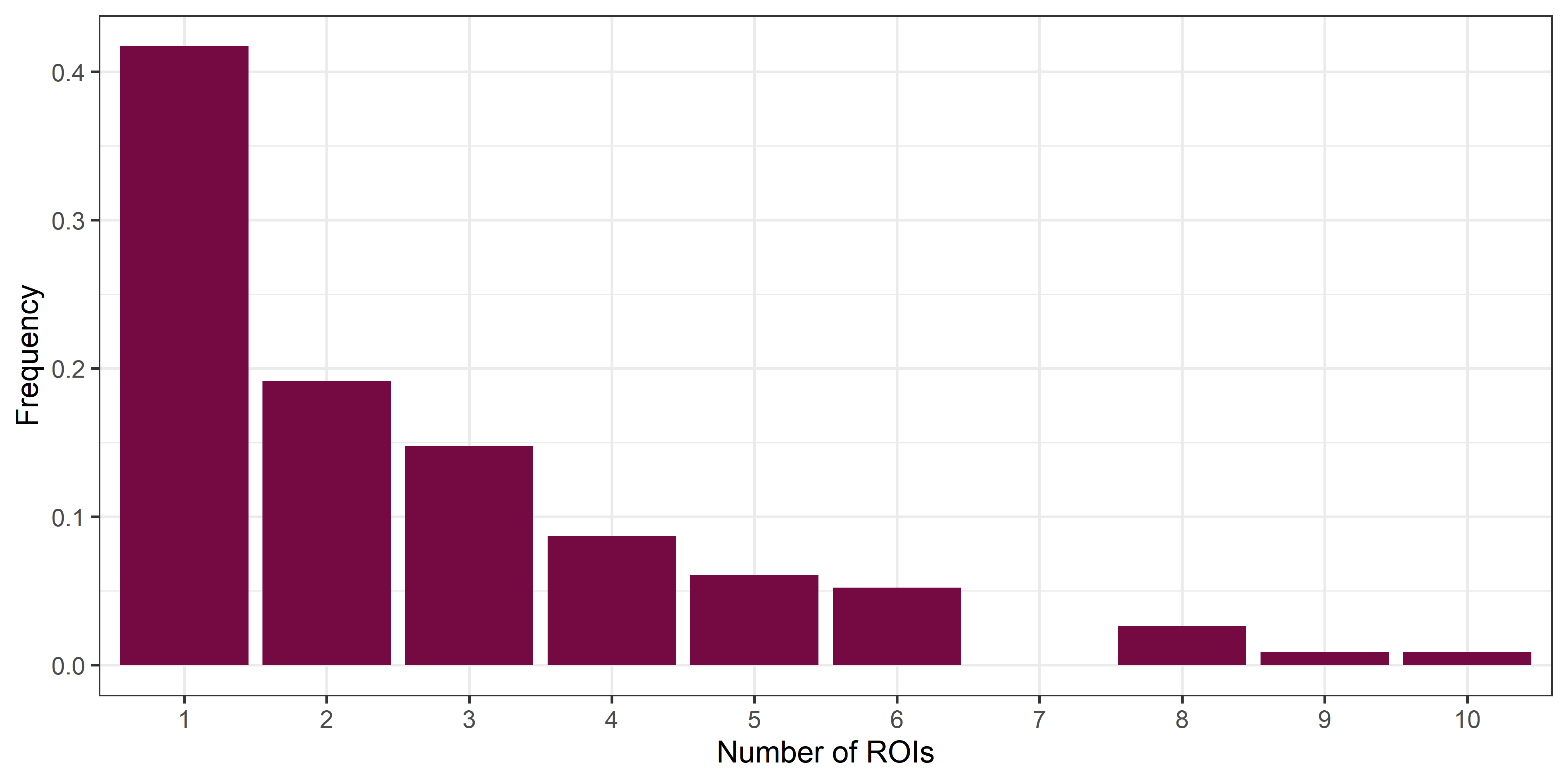}
\caption{Characterization of the cohort with respect to ROI number}
\label{fig:nROI}
\end{figure}

The clinical characteristics of the study population were consistent with a typical lung cancer cohort, with the majority of patients being male and diagnosed in an advanced stage of the disease (according to the TNM classification). 
The patients differed in the number of contoured regions of interest, ranging from one to ten (see Figure \ref{fig:nROI}). In nearly 60\% cases, multiple ROIs were present. 

\subsection{Inter-ROI heterogeneity related to early metastasis}

\begin{figure*}[!t]
\centering
\begin{subfigure}[t]{0.45\textwidth}
\includegraphics[width=0.9\textwidth]{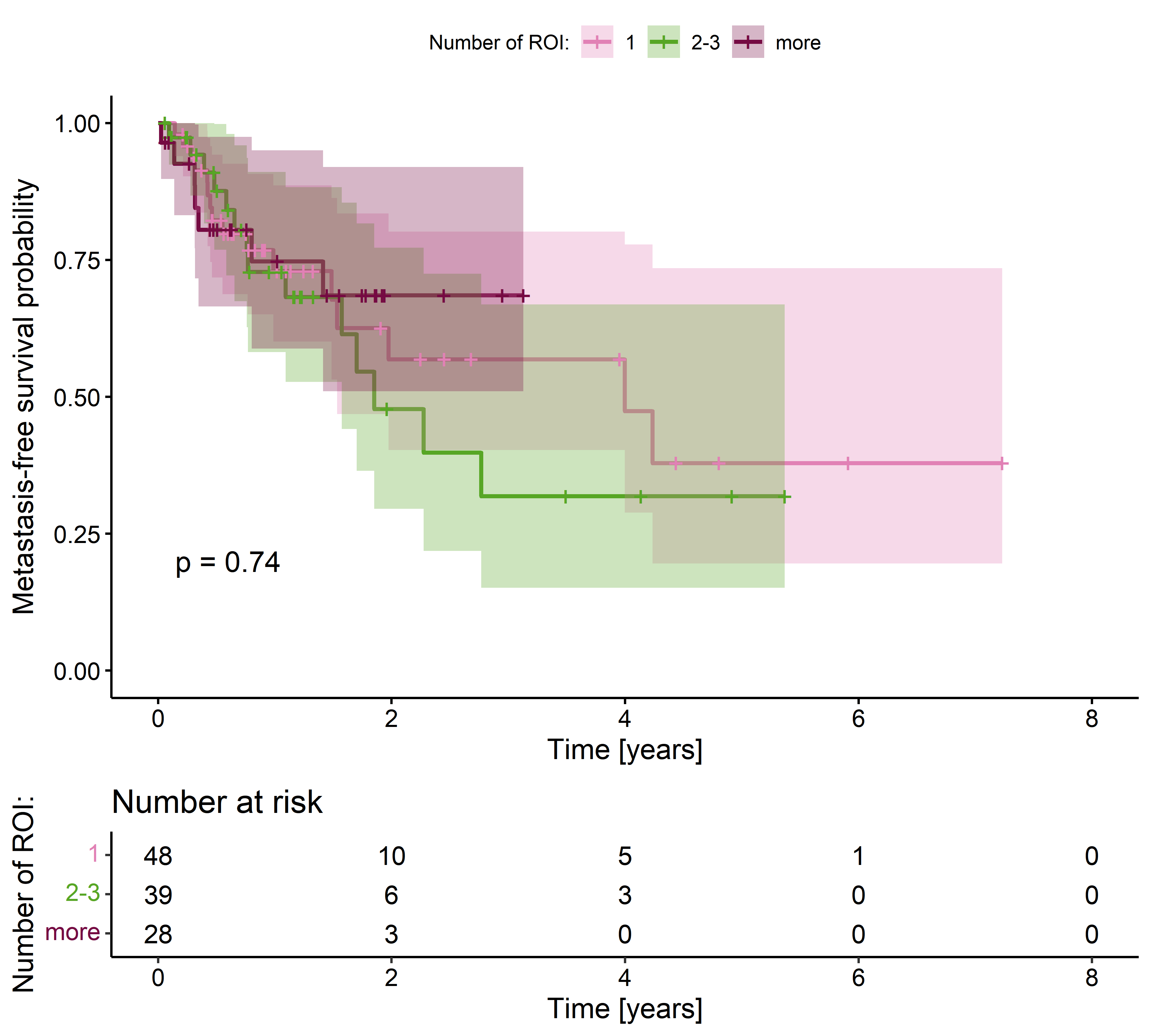}
\caption{}
\label{fig:KMnROI}
\end{subfigure}
\begin{subfigure}[t]{0.45\textwidth}
\includegraphics[width=0.95\textwidth]{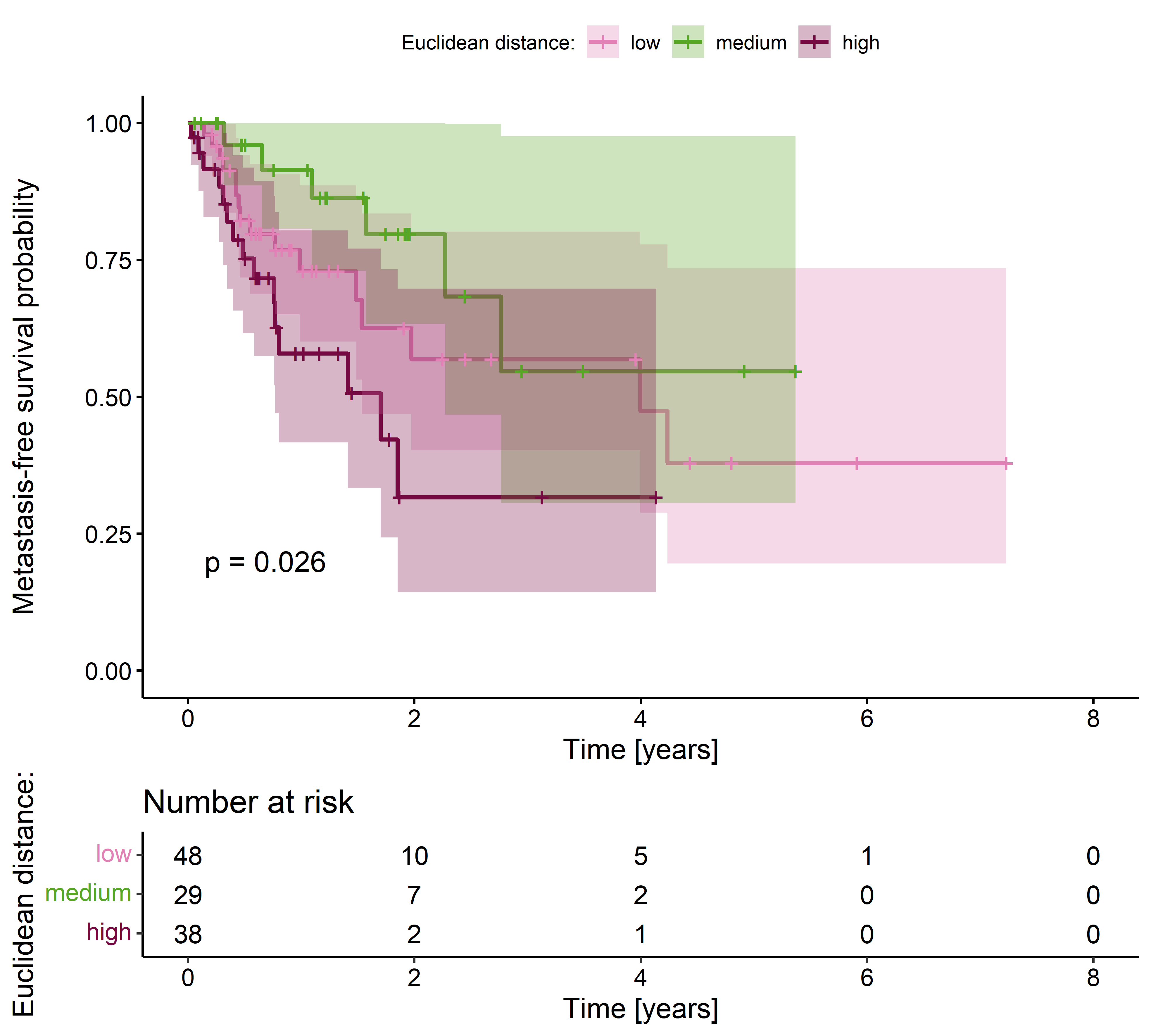}
\caption{}
\label{fig:KMeuclid}
\end{subfigure}
\caption{Predictive value of the inter-ROI heterogeneity. (a) Kaplan-Meier plot of metastasis-free survival for patients divided according to the number of ROIs. (b) Kaplan-Meier plot of metastasis-free survival for patients divided according to the heterogeneity index of ROIs (euclidean distance, PET resolution)}
\label{fig:result}
\end{figure*}

In the analysed cohort, there was no significant difference in metastasis-free survival probability between the groups based on ROI number (Figure \ref{fig:KMnROI}). In contrast, for the PET\_CT dataset, the two subgroups with more than one ROI differed significantly for some of the diversity indices, with higher heterogeneity related to early metastasis. The lowest p-value equal to 0.026 was obtained for the Euclidean distance (Figure \ref{fig:KMeuclid}). It shows that the diversity indices can be used as variables in a survival model for MFS prediction. However, the Kaplan-Meier curve for the low-heterogeneity group was positioned between the other two, suggesting that the inclusion of other features is necessary for the generality of the model.

\subsection{Comparison of approaches to handling multiple ROIs and survival models}

\begin{figure*}
\centering

\sbox{\bigimage}{%
  \begin{subfigure}[b]{.6\textwidth}
  \centering
  \includegraphics[width=\textwidth]{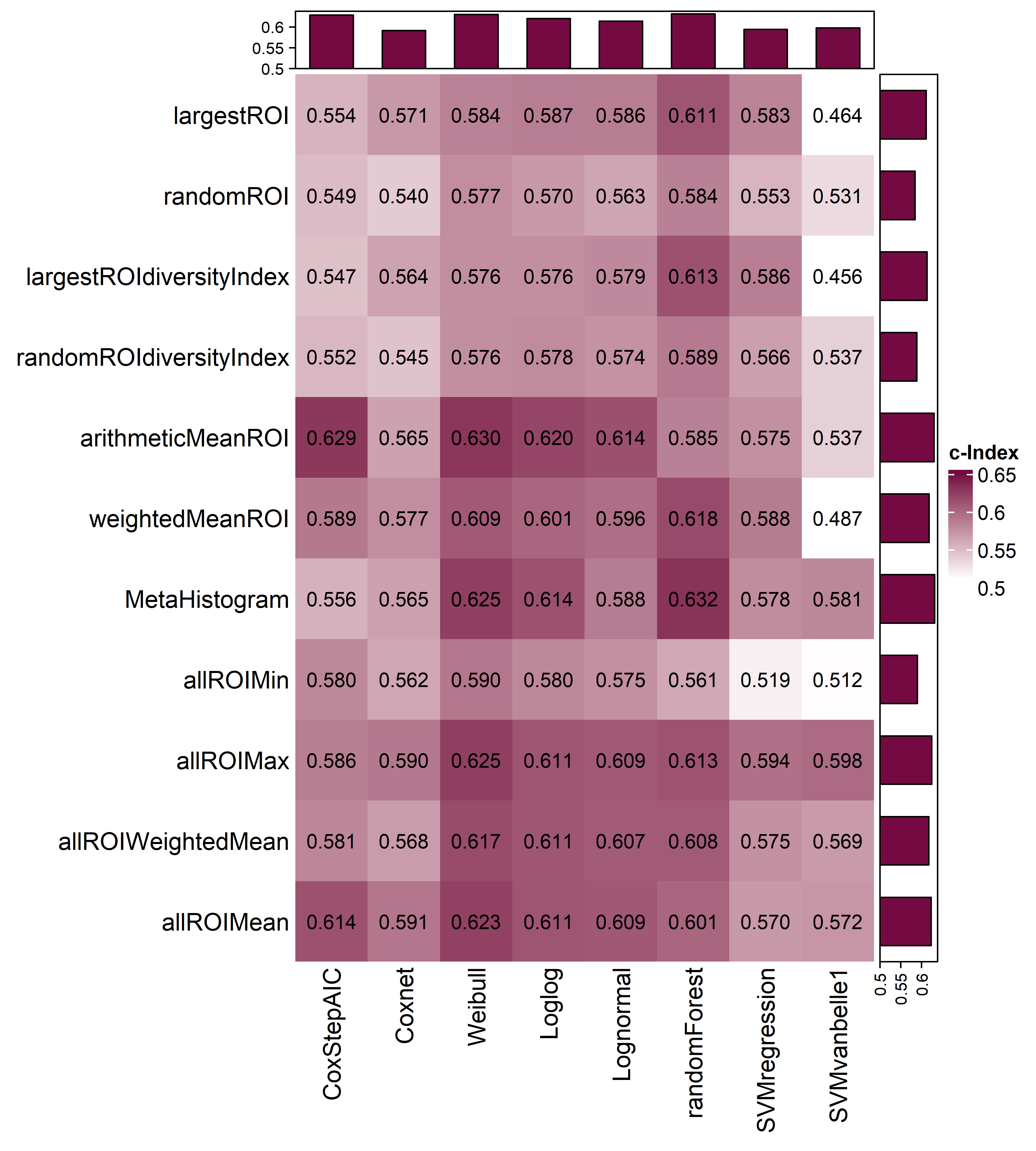}
  \caption{}
\label{fig:benchmarkPET}
  \vspace{0pt}% reference point at the very bottom
  \end{subfigure}%
}

\usebox{\bigimage}\hfill
\begin{minipage}[b][\ht\bigimage][s]{.4\textwidth}
  \begin{subfigure}{\textwidth}
  \centering
  \includegraphics[width=\textwidth]{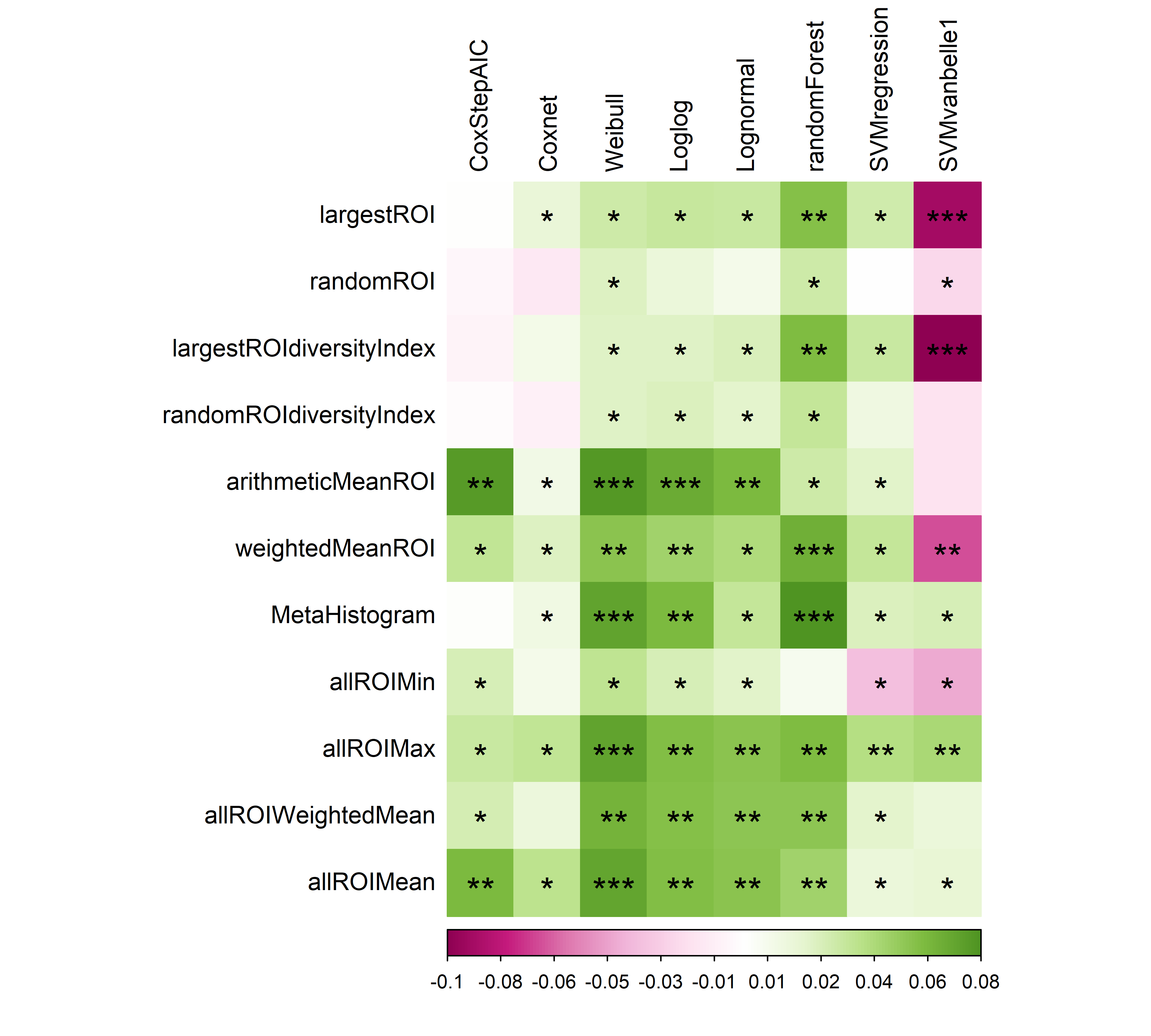}
  \caption{}
  \label{fig:cohendPET}
  \end{subfigure}%
  \vfill
  \begin{subfigure}{\textwidth}
  \centering
  \includegraphics[width=\textwidth]{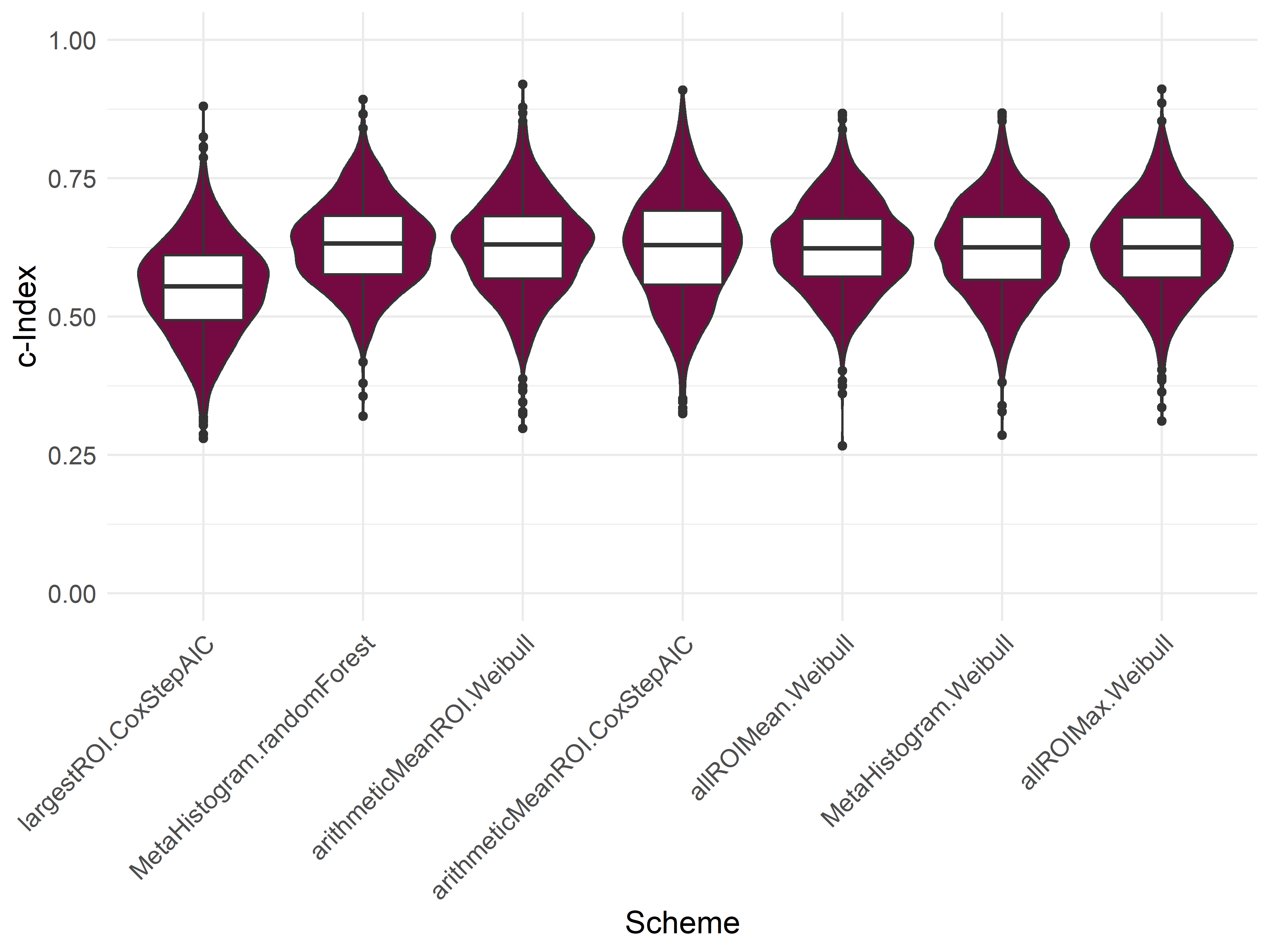}
  \caption{}
  \label{fig:MCCV_boxPET}
  \end{subfigure}

  \vspace{0pt}% reference point at the very bottom
\end{minipage}
\qquad
\caption{Results of the comparison for the PET dataset. (a) Median c-Indices of all the tested schemes. The top and right annotations show the best median c-Indices for the models and ROI handling methods, respectively. (b) Comparison of the schemes with the reference model (largestROI+CoxStepAIC). Colours represent differences between median c-Indices, and stars show Cohen's d effect size,  ( ) meaning negligible, (*) --- small, (**) --- medium, and (***) --- large. (c) c-Indices of the selected schemes across 1000 iterations.}
\label{fig:resultPET}
\end{figure*}

For the PET dataset, the model quality varied both between models and ROI handling approaches (Figure \ref{fig:benchmarkPET}), with median c-Indices between 0.456 for the combination\textit{ largestROIdiversityIndex + SVMvanbelle1 }and 0.632 for \textit{MetaHistogram + randomForest}. For the reference scheme, \textit{largestROI + CoxStepAIC}, the median c-Index was 0.554. 7 schemes performed better with large effect size, 20 performed better with medium effect size, 37 performed better with small effect size, 3 performed worse with small effect size, 1 performed worse with medium effect size, and 2 performed worse with large effect size. For the rest, the effect size was negligible (Figure \ref{fig:cohendPET}). 

 \textit{randomForest}, \textit{Loglog}, and \textit{Weibull} generally yielded high c-Indices. Interestingly, for the \textit{largestROI} method, all but one model were better than the reference (\textit{CoxStepAIC}). Although adding the diversity indices improved the \textit{largestROI} method only for two models (\textit{randomForest} and \textit{SVMregression}), it was beneficial for the \textit{randomROI} method, improving the performance of all but one model.

The proposed methods of handling multiple ROIs allowed for improving the predictive ability of the models. Unsurprisingly, \textit{randomROI} worked worse than the \textit{largestROI} approach for every model but \textit{SVMvanbelle1}. Among the ROI aggregation methods, \textit{arithmeticMeanROI} achieved the highest overall median c-indexes --- 0.629 for \textit{CoxStepAIC} and 0.630 for \textit{Weibull}. Although not as effective, \textit{weightedMeanROI} method was more consistent across different models, outperforming \textit{largestROI} for all of them. Risk aggregation, with the exception of \textit{allROIMin}, also yielded good results, particularly for the Weibull model --- 0.623 for \textit{allROIMean} and 0.625 for \textit{allROIMax}. The latter method achieved higher c-Indices than \textit{largestROI} for all tested models. 

The schemes also differed in the distribution of c-Indices across the cross-validation iterations (Figure \ref{fig:MCCV_boxPET}). The minimum c-Indices were between 0.090 for the scheme \textit{weightedMeanROI + SVMvanbelle1} and 0.364 for \textit{allROIMax + randomForest}, while the maximum c-Indices ranged from 0.733 for \textit{largestROI + SVMvanbelle1} to 0.941 for \textit{randomROIdiversityIndex + SVMregression}. Since c-Indices below 0.5 indicate a performance worse than random, such values for the test set likely stem from overfitting. Schemes \textit{MetaHistogram + randomForest} \textit{allROIMean + Weibull}, \textit{weightedMean + randomForest}, and \textit{allROIMax + Weibull} were most robust to overfitting, as all of them had c-Index $<$ 0.5 in fewer than 60 out of 1000 iterations. A complete summary of the results is given in the Supplementary Table 2.

\begin{figure*}
\centering

\sbox{\bigimage}{%
  \begin{subfigure}[b]{.6\textwidth}
  \centering
  \includegraphics[width=\textwidth]{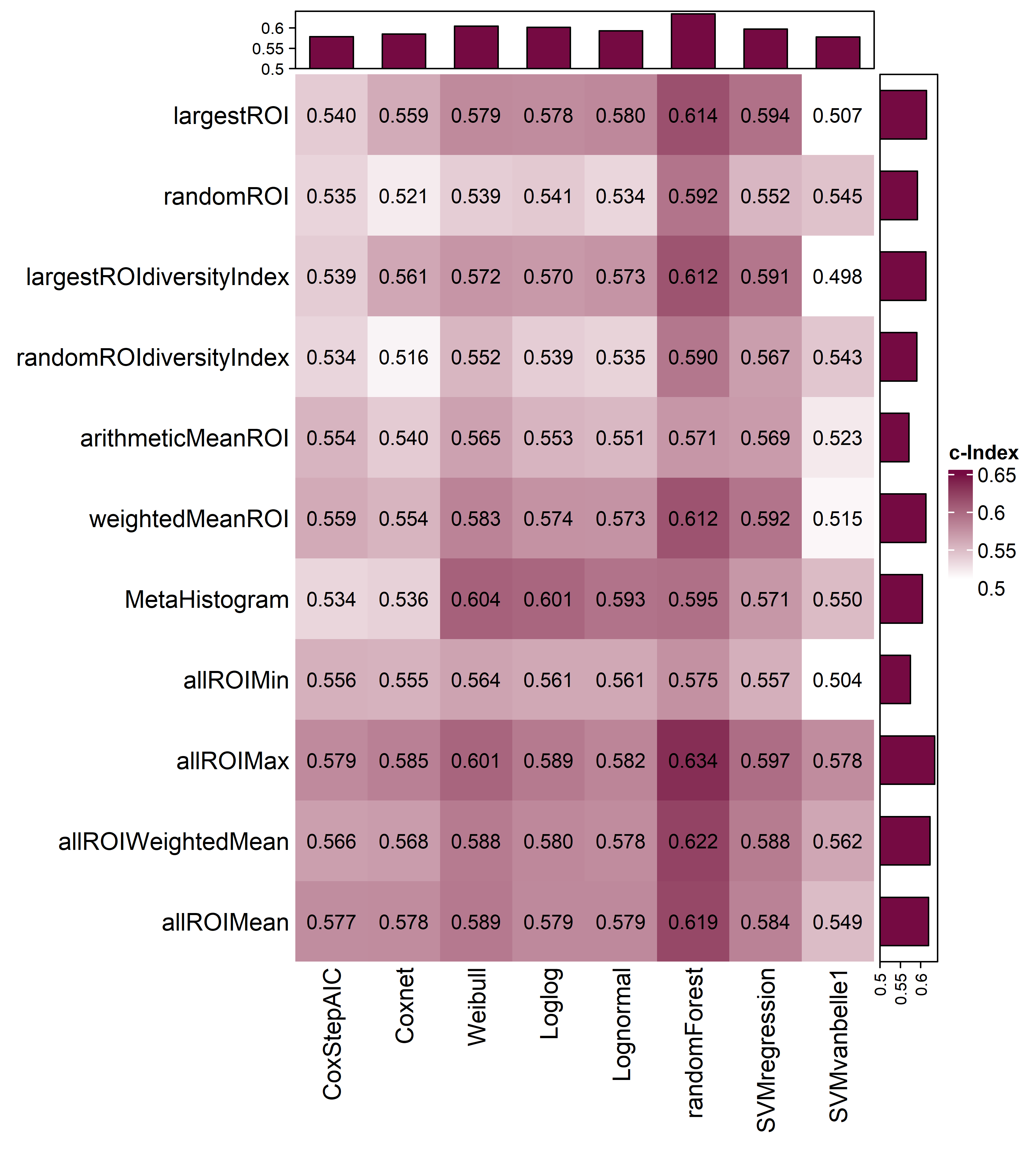}
  \caption{}
\label{fig:benchmarkPETCT}
  \vspace{0pt}% reference point at the very bottom
  \end{subfigure}%
}

\usebox{\bigimage}\hfill
\begin{minipage}[b][\ht\bigimage][s]{.4\textwidth}
  \begin{subfigure}{\textwidth}
  \centering
  \includegraphics[width=\textwidth]{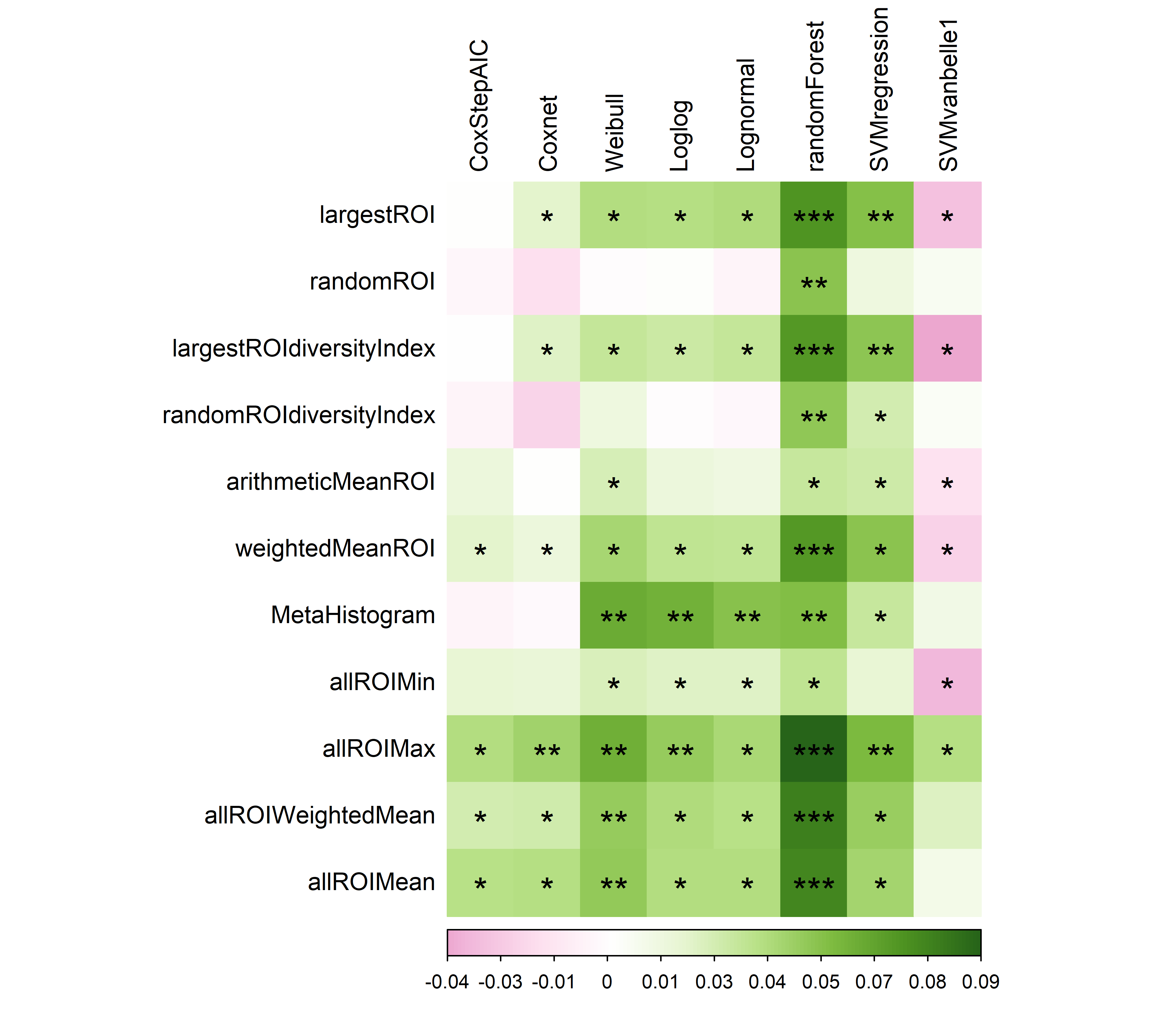}
  \caption{}
  \label{fig:cohendPETCT}
  \end{subfigure}%
  \vfill
  \begin{subfigure}{\textwidth}
  \centering
  \includegraphics[width=\textwidth]{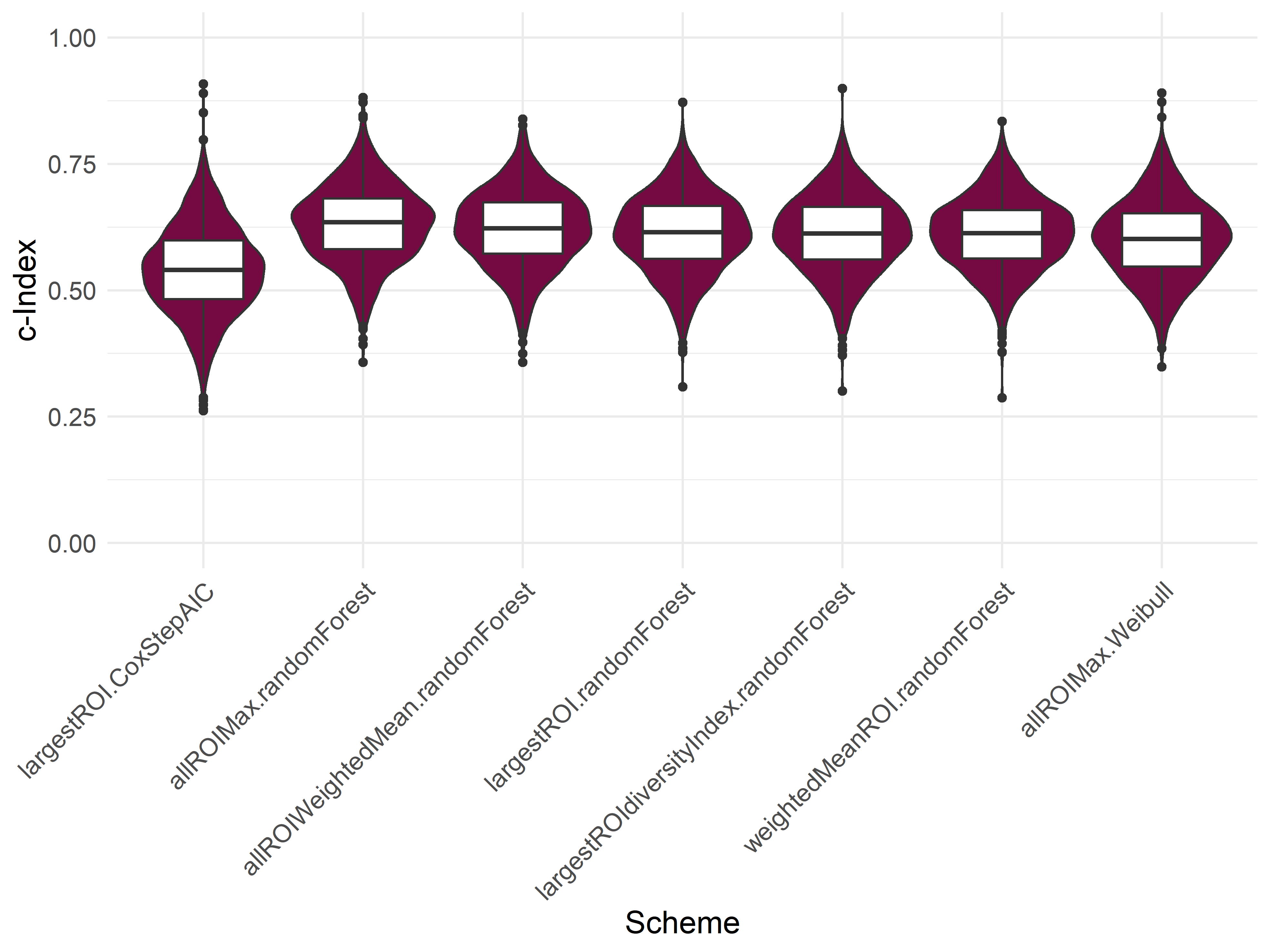}
  \caption{}
  \label{fig:MCCV_boxPETCT}
  \end{subfigure}

  \vspace{0pt}% reference point at the very bottom
\end{minipage}
\qquad
\caption{Results of the comparison for the PET\_CT dataset. (a) Median c-Indices of all the tested schemes. The top and right annotations show the best median c-Indices for the models and ROI handling methods, respectively. (b) Comparison of the schemes with the reference model (largestROI+CoxStepAIC). Colours represent differences between median c-Indices, and stars show Cohen's d effect size,  ( ) meaning negligible, (*) --- small, (**) --- medium, and (***) --- large. (c) c-Indices of the selected schemes across 1000 iterations. }
\label{fig:resultPETCT}
\end{figure*}

For the PET\_CT dataset, similar tendencies could be observed with large differences between schemes (Figure \ref{fig:benchmarkPETCT}); the median c-Indices varied between 0.498 for the combination \textit{largestROIdiversityIndex + SVMvanbelle1} and 0.634 for \textit{allROIMax + randomForest}. For the reference scheme, \textit{largestROI + CoxStepAIC}, the median c-Index was 0.540. 6 schemes performed better with large effect size, 14 performed better with medium effect size, 35 performed better with small effect size, and 5 performed worse with small effect size. For the rest, the effect size was negligible (Figure \ref{fig:cohendPETCT}). 

The \textit{randomForest} model consistently outperformed others regardless of the ROI handling approach, with the sole exception of the \textit{MetaHistogram} approach. For the \textit{largestROI} method, all but one model was better than the reference (\textit{CoxStepAIC}). Once again, the worst performing model for most ROI handling methods was \textit{SVMvanbelle1}.

Also for this dataset, the results obtained for the \textit{largestROI} method could be improved by incorporating multiple ROIs. Risk aggregation methods seemed to be more suitable --- \textit{allROIMax} achieved better c-Indices for all models, \textit{allROIMean} and \textit{allROIWeightedMean} performed similarly or better than \textit{largestROI}. 

The schemes displayed different robustness levels observed in the distribution of c-Indices across the cross-validation iterations (Figure \ref{fig:MCCV_boxPETCT}). The minimum c-Indices were between 0.062 for the scheme \textit{weightedMeanROI + SVMvanbelle1} and 0.357 for \textit{allROIMax + randomForest}, while the maximum c-Indices ranged from 0.762 for \textit{weightedMeanROI + SVMvanbelle1} to 0.922 for \textit{allROIMax + SVMregression}. Schemes \textit{allROIMax + randomForest}, \textit{allROIWeightedMean + randomForest}, \textit{allROIMean + randomForest}, and  \textit{weightedMeanROI + randomForest} were most robust to overfitting, achieving c-Indices lower than 0.5 in 50, 60, 61 and 61 iterations, respectively. A complete summary of the results is given in the Supplementary Table 3.

\section{Discussion}
Radiomics is gaining popularity as a source of biomarkers, both in diagnostic and prognostic settings \cite{TAGLIAFICO2020,Wu2018,Liu2019,DAmico2020,Homayoun2022}. The extraction of radiomic features requires a well-defined region of interest (ROI), which in cancer research is often centered on the primary tumor. However, many cancers are already locally or regionally disseminated at the time of diagnosis. For instance, between 2011 and 2020 in the United States, only 23.3\% of lung cancer patients were diagnosed at a localized stage, while 21.1\% were at a regional stage, and 48.3\% at a distant stage \cite{SEER}. In regionally advanced cancers, the number of potential ROIs often exceeds one, encompassing lymph nodes and secondary lesions in addition to the primary tumour. These additional regions may harbor prognostically relevant information, but their integration into predictive models is methodologically challenging since their number and nature vary between patients. As a result, they are often overlooked.

This work seeks to address this gap by proposing strategies for integrating radiomic information from multiple distinct lesions into survival models. Our aim is primarily to establish a framework for fully utilising the available data and compare its effectiveness to a primary-tumour-only strategy, rather than to answer a specific clinical question. To explore the possibility of incorporating multiple ROIs into survival-type models, we collected a cohort of 115 non-metastatic non-small cell lung cancer patients, with between one and ten contoured ROIs. We used radiomic features extracted from the ROIs to predict metastasis-free survival, employing different regression models and proposing various strategies of multiple ROI handling. Additionally, we evaluated the relation between inter-ROI heterogeneity and time-to-metastasis.

We have shown that for patients with more than one ROI, their number alone was not significantly related to the metastasis-free survival probability (see Figure \ref{fig:KMnROI}). However, for the PET dataset, the groups created by dividing patients by heterogeneity indices differed in MFS --- higher inter-ROI heterogeneity was related to a higher risk of metastasis (Figure \ref{fig:KMeuclid}). Therefore, heterogeneity indices, synthetic features combining information from all the patient's ROIs, can be used as prognostic variables in the models. Still, the trend is disturbed, as the Kaplan-Meier curve for a group of patients with only one ROI, and consequently low heterogeneity, was between the two others. For a more generally useful model, the heterogeneity indices must be augmented with other radiomic features.

The prediction quality, evaluated in terms of Harrell's c-Index, was highly dependent on the used survival model (Figures \ref{fig:resultPET}, \ref{fig:resultPETCT}). While the classic approach, consisting of the Cox proportional hazards regression with stepwise selection based on AIC performed quite well, other models achieved better predictions --- random survival forest achieved higher c-Index for each ROI handling method. Ensemble models, such as model-based boosting and the random forest, deserve attention for their performance. They model the data well without overfitting, as seen on the test data, where they rarely yielded c-Indices below 0.5 (Supplementary Tables 2 and 3). 

The ROI aggregation methods performed better for the PET dataset. Adding diversity indices to radiomic features allowed for improving some largest-ROI models, and many randomROI models. Even better results were obtained for the \textit{arithmeticMeanROI} and \textit{weightedMeanROI}, which outperformed the \textit{largestROI} method for most models. This confirms the benefit of incorporating information from all lesions, especially since in ROI aggregation methods the improvement cannot be explained by a larger data set used for training.

Very good overall results for both datasets were achieved by the risk aggregation methods except for allROIMin. An intuitive explanation of this poor performance can be found in the inter-ROI heterogeneity analysis. Patients with highly diverse ROIs will likely be assigned both very high and very low risks. After risk aggregation, they will be given the lowest value, contrary to what is presented in Figure \ref{fig:KMeuclid}, where the high-heterogeneity group exhibits more and earlier events than the other two. The opposite strategy implemented in the allROIMax method is consistent with the findings from the heterogeneity analysis, resulting in the highest c-Indices of all the methods.

The \textit{MetaHistogram} method, introduced by Zhao and coworkers \cite{Zhao2022}, improved the model performance compared to the largest ROI particularly in the PET dataset. However, since the endpoint of this study was related to metastasis, our cohort was limited to cases where cancer was regionally advanced at most. Consequently, the number of ROIs was relatively smaller than in the original work introducing the concept –-- in fact, slightly over 40\% of patients had only one contoured lesion. In such cases, the constructed “meta histogram” consists of only one bin, which makes some of the extracted features uninformative. 

While radiomics is well-established as a powerful approach for extracting prognostic and predictive biomarkers, new methods are constantly being developed to enhance the performance of survival models. For instance, incorporating autoencoder-based features improved the c-Index for survival prediction in non-small cell lung cancer from 0.58 (radiomics-only) to 0.63 \cite{Ferretti2025}. Similarly, Mazher et al. employed a 3D Latent CNN and 3D Deep Regressor, achieving a c-Index of 0.63 for brain tumor survival prediction, compared to 0.57 using only traditional radiomic features \cite{Mazher2023}. In another approach, delta-radiomics applied to non-small cell lung cancer patients treated with immune checkpoint inhibitors increased the c-Index from 0.62 (radiomics-only) to 0.68 \cite{Cousin2023}. Importantly, as our approach focuses on ROI choice, it can be combined with other innovative methods such as deep or delta radiomics for potentially even better results.

This study is not without limitations, the greatest being only one, relatively small cohort used for comparing the strategies. A search was conducted for an external imaging dataset to validate the findings. While there are studies providing multiple nodules per patient \cite{Armato2011} or survival data \cite{Aerts2014}, to the authors' best knowledge no public dataset exists with both available. It should however be noted, that the aim of the present study is not to present a solution to a particular clinical problem (in this case the prediction of metastasis risk) but to investigate whether incorporating information from multiple lesions can improve model performance.

% Also, different possible ROIs have been explored \cite{Xu2022,Feng2022,Shan2019,Chen2022B,Chen2018,Zhou2020,Han2020,Dammak2024,Xin2023,Wang2024,Peng2024,Zhang2023,Zhang2024,Nie2024,Hou2024}, usually related to either narrowing or extending the tumour area, for instance tumour core, tumour margin, peritumoural area etc. Including such regions in a machine learning model is relatively straightforward, since they can be identified for every patient and accordingly labelled, introducing no ambiguity. In our work, we understand multiple regions of interest as independent uptakes in the PET images, as opposed to the elsewhere adopted idea based on different ways of contouring or defining new types of ROI. In one study also involving multiple actual lesions, the method of extracting multi-lesion radiomic features was used for a classification problem \cite{Zhao2022}. Nevertheless, we included the described strategy (\textit{MetaHistogram}) in our comparison demonstrating that for certain datasets it can indeed improve the performance of survival models. 

\section{Conclusion} 

Incorporating other lesions in addition to the primary tumor, consistently improved the performance of survival models across various regression approaches. The aggregation strategies proposed in this work offer a robust solution for handling multiple radiomics ROIs, but also have broader applicability. Using all available feature vectors in a structurally heterogeneous dataset to fit survival models offers a promising direction, not only for imaging studies but also for molecular research, particularly in spatial or single-cell analysis.

\section*{Acknowledgments}
This work was supported by the Polish National Science Centre, grant number: UMO-2020/37/B/ST6/01959.%, Medical Research Agency, grant number: 07/010/ABB23/1037 and Silesian University of Technology statutory research funds. 
Calculations were performed on the Ziemowit computer cluster in the Laboratory of Bioinformatics and Computational Biology created in the EU Innovative Economy Programme POIG.02.01.00-00-166/08 and expanded in the POIG.02.03.01-00-040/13 project.

\section*{Competing interests}
The authors declare that they have no known competing financial interests or personal relationships that could have appeared to influence the work reported in this paper.
 
\section*{Data and code availability}
The data used in this work and source code are available upon reasonable request.

\section*{Author contributions}
% CRediT - Imię Nazwisko: role 
\textbf{Agata Ma\l gorzata Wilk}: Conceptualization, Methodology, Software, Formal Analysis, Investigation, Visualization, Writing --- Original Draft. \textbf{Andrzej Swierniak}: Conceptualization, Supervision, Project administration, Funding acquisition, Writing --- Review \& editing. \textbf{Andrea d'Amico}: Investigation. \textbf{Rafa\l ~Suwinski}: Resources. \textbf{Krzysztof Fujarewicz}: Project administration, Supervision, Writing --- Review \& editing.  \textbf{Damian Borys}: Data curation, Software, Investigation, Visualization, Writing --- Review \& editing.

\bibliographystyle{elsarticle-num-names.bst}
\bibliography{biblio1}

\end{document}